# DESIGN PRINCIPLES DEVELOPED THROUGH USER-CENTERED AND SOCIO-TECHNICAL METHODS IMPROVE CLINICIAN SATISFCATION, SPEED, AND CONFIDENCE IN PHARMACOGENOMIC CLINICAL DECISION SUPPORT

Timothy M. Herr, PhD[1]; Therese A. Nelson, AM, LSW[2]; Luke V. Rasmussen, MS[1]; Yinan Zheng, PhD[1]; Nicola Lancki, MPH[1]; Justin B. Starren, MD, PhD, FACMI[1]

[1]Department of Preventive Medicine,

Northwestern University Feinberg School of Medicine, Chicago, IL

[2]NUCATS Institute,

Northwestern University Feinberg School of Medicine, Chicago, IL

## Abstract

OBJECTIVE: To design and evaluate new pharmacogenomic (PGx) clinical decision support (CDS) alerts, built to adhere to PGx CDS design principles developed through socio-technical approaches.

MATERIALS AND METHODS: Based on previously identified design principles, we created 11 new PGx CDS alert designs and developed an interactive web application containing realistic clinical scenarios and user workflows that mimicked a real-world EHR system. We recruited General Internal Medicine and Cardiology clinicians from Northwestern Medicine and recorded their interactions with the original and new designs. We measured clinician response, satisfaction, speed, and confidence through questionnaires and analysis of the recordings.

RESULTS: The study included 12 clinicians. Participants were significantly more satisfied (p=0.0000001), faster (p=0.009), and more confident (p<.05) with the new designs than the original ones. The study lacked statistical power to determine whether prescribing accuracy was improved, but participants were no less accurate, and clinical actions were more concordant with alert interactions (p=0.004) with the new designs. We found a significant learning curve associated with the original designs, which was eliminated with the new designs.

DISCUSSION: This study successfully demonstrates that socio-technical and user-centered design techniques can improve PGx CDS alert designs. Best practices for PGx CDS design are limited in the literature, with few effectiveness studies available. These results can help guide future PGx CDS implementations to be more clinician friendly and less time-consuming.

CONCLUSION: The results of this study support the PGx CDS design principles we proposed in previous work. As a next step, the new designs should be implemented in a live setting for further validation.

**Background and Significance**

Traditional approaches to drug prescribing have inefficiencies that create a substantial burden on the U.S. healthcare system. Some patients experience expensive and life-altering adverse drug events and adverse drug reactions, while others experience a lengthy trial-and-error process to find therapeutic drugs and dosages.[1] Precision medicine has the potential to mitigate these issues by tailoring drug prescribing more accurately for the individual patient.[2] Pharmacogenomics (PGx) has shown promise as a route for meaningful implementation of precision medicine in a variety of clinical scenarios.[3-7] However, clinicians are not currently well educated on the benefits of PGx, or on methods to implement it effectively in treating patients.[8,9] Additionally, the rapidly changing field of genomics makes it hard for any one individual to keep up with the shifting science and associated clinical recommendations. Clinical decision support (CDS) systems have the ability to close this gap by conveying useful PGx knowledge to prescribers at the point of care,[10] without requiring time-consuming continuing medical education that most clinicians find burdensome.[11]

To-date, PGx CDS systems have shown mixed results in their level of effectiveness and clinician-acceptance. The wide variation in the design and outcomes of first-generation PGx CDS systems confirms that there are no current, widely accepted best practices for how to design alerts to be as effective and useful for clinicians as possible. We previously examined the alerts implemented at Northwestern Medicine (NM) and found generally low acceptance and compliance rates with PGx recommendations.[12] Other sites in the Electronic Medical Records and Genomics (eMERGE) Network showed a wide range of effectiveness and design choices, though with some emerging consensus around the use of post-genetic test, interruptive alerts based on Clinical Pharmacogenetics Implementation Consortium (CPIC) recommendations integrated into the EHR.[13] Non-eMERGE sites have taken a variety of different approaches to their implementations, as well. The Genomic Prescribing System at the University of Chicago is a standalone system that has been shown to affect prescriber decision-making using "traffic light" recommendations,[14,15] while St. Jude's and the NIH Clinical Center developed interruptive alerts, with substantially different clinician compliance rates.[16,17]

Although a number of sites have opted for interruptive alerts, there remains wide variation in the design and function of those alerts. For PGx CDS to reach its potential, we must establish best practices and guidelines for alert design to ensure that knowledge is conveyed to prescribers in a meaningful way that will actually improve patient care and outcomes. Non-PGx CDS has a broad base of research to draw upon,[18-21] but low levels of provider education in PGx raises the possibility that information needs for PGx CDS may be substantially different from areas where clinicians have been extensively trained, such as drug-drug and drug-allergy interactions.

Socio-technical design methods have the potential to help identify clinician needs and best practices for PGx CDS. These methods emphasize inclusion of the user in the design process in order to best evaluate actual needs. They also utilize repeated evaluation of, and iteration upon, designs.[22-24] In prior work, we applied these techniques and identified several potential principles for PGx CDS alert design.[25] In this study, we build upon that work and propose and evaluate a series of updated PGx CDS designs based on those principles, in comparison to the previous committee-developed designs at NM. We suggest that the designs based on socio-technically developed principles will lead to better response by clinicians on a variety of metrics. If effective, then these principles can be adopted more broadly in the field to ensure PGx CDS reaches its potential for bringing precision medicine to fruition.

**Objective**

This study sought to compare clinician response to PGx CDS (including action taken, time taken, confidence, and overall satisfaction) for alert designs built with two different approaches. The first approach was a set of committee-developed designs, established through informal user input and previously implemented at Northwestern Medicine. The second approach was a set of revised designs, based on the results of formal socio-technical design work described in a previous paper.[25]

**Materials and Methods**

Overview

Participants in this study took part in a user test consisting of a series of 24 interactive simulations based on fictionalized, but realistic, clinical scenarios involving PGx CDS and were asked to engage in a think-aloud protocol. Simulations were designed to mimic familiar EHR workflows. Twelve scenarios used existing PGx CDS alert designs, developed in earlier phases of NU's eMERGE-PGx project ("original designs"). Twelve scenarios used new PGx CDS alert designs, developed via socio-technical design methods ("new designs"). After each scenario, participants were asked to rate their confidence in their actions and explain their reasoning if they chose to ignore an alert. After completing all 24 scenarios, participants completed a short questionnaire assessing their preferences between the different designs. All testing sessions included screen and audio recording for later analysis. The Northwestern IRB approved all aspects of this study.

Study Development

*Alert Development*

We first developed a series of new PGx CDS design prototypes. These designs were based on the results of prior research and on general human-computer interaction and user interface design principles applicable to clinical decision support.[18-21,25] New design prototypes were initially developed by TMH and then iteratively refined based on group feedback (JBS, LVR, TAM).  The design principles established from prior research, and how they were addressed in the new designs, are detailed in Table 1.  Some of these principles and approaches are consistent with those used in traditional CDS, but several are unique to PGx CDS, including the use of stronger wording in recommendations, the use of phenotypes abstracted from the underlying genotype data, and emphasis on adaptability for learning effects. Figure 1 provides an example comparing an original design to a new design.

Table 1 – Pharmacogenomic Clinical Decision Support Design Principles Used

| Design Principle[a] | Addressed By… |
|---|---|
| Be Specific and Actionable | <ul><li>A large, bold, colored banner stating exactly what the alert is for (e.g., "MEDICATION CHANGE RECOMMENDED")</li><li>An exact recommended alternative medication and dose instead of a list of "alternatives to consider"</li><li>Accepting an alert or requesting educational materials automatically updates the user's order without additional clicks</li><li>Desired additional actions are clearly laid out with buttons to confirm their completion</li></ul> |
| Be Brief | <ul><li>Longer verbiage is de-emphasized in favor of color-coded medication names that make it clear what is recommended and what is not</li><li>Educational materials are available by a link and longer justifications are placed lower in the alert</li></ul> |
| Display Phenotypes not Genotypes | <ul><li>Instead of genetic results like "CYP2C19 *2/*2," alerts contain phenotypes like "Clopidogrel Poor Metabolizer"</li><li>All genetic results are available via links to laboratory results for those who are interested</li></ul> |
| Rely on Sources Clinicians Already Trust | <ul><li>Alerts mention Northwestern Medicine or the Northwestern Pharmacy & Therapeutics committee</li></ul> |
| Be Adaptable to Learning Effects | <ul><li>Some alerts appear in summary form that can be expanded for more details</li><li>All alerts are designed to be easily skimmed by those with more experience, with details and links for those that need more information</li></ul> |

[a] Design principles derived from previous work.[25]

Figure 1 – PGx CDS Design Comparison

*Original Design:*

*New Design, Brief:*

*New Design, Expanded:*

> **BestPractice Advisory - Zztest, Intermediate**
>
> ⚠ **Drug-Gene Interaction Alert** - Northwestern Medicine and Clinical Pharmacogenetics Implementation Consortium (CPIC) Recommendation
>
> ## MEDICATION CHANGE RECOMMENDED
>
> *Currently Selected Medication*
>
> clopidogrel, 75mg
>
> *Recommended Alternative*
>
> prasugrel, 10mg
>
> *Clinical Indication*
>
> ↰ CLOPIDOGREL POOR METABOLIZER
>
> *Additional Actions*
>
> 1. Discuss Results with Patient        2. Order Educational Materials for Patient
>
> [ Acknowledge Discussion ]            [ Add AVS Materials to Order ]
>
> *Justification*
>
> Genetic factors are known to affect drug metabolism. Based on genetic test results on file, clopidogrel may be less effective for this patient and increase risk for adverse cardiovascular events.
>
> Evidence Level: Moderate
> ↰ View CPIC Reference Materials
> ↰ View Fact Sheet
>
> © 2019 Epic Systems Corporation. Used with permission.        ✓ Accept        Dismiss

The new alerts were developed for the same drug-gene interactions (DGIs) and clinical scenarios as in previous eMERGE-PGx work, but with updated UIs and workflows. Table 2 details each of the eleven distinct alerts that were developed.

Table 2 – Pharmacogenomic Clinical Decision Support Alerts Developed

| Drug | Active/Passive | Predicted Phenotype | Medication Status |
|---|---|---|---|
| Clopidogrel | Active | Intermediate Metabolizer | Actively Prescribing |
| | | Poor Metabolizer | Actively Prescribing |
| | Passive | Intermediate Metabolizer | Not on Medication |
| | | Intermediate Metabolizer | On Medication |
| | | Poor Metabolizer | Not on Medication |
| | | Poor Metabolizer | On Medication |
| Simvastatin | Active | Intermediate Activity | Actively Prescribing |
| | | Low Activity | Actively Prescribing |
| | Passive | Intermediate Activity | Not on Medication |
| | | Low Activity | On Medication |
| Warfarin | Active | Sensitive | Actively Prescribing |

Active alerts are defined as interruptive alerts that appear when a clinician chooses one of the affected drugs during order entry and the patient has a relevant genetic result on file. Passive alerts are defined as non-interruptive alerts that appear in the "Best Practice Advisories" tab of the patient's EHR chart when the patient has a relevant genetic test result on file. Passive alerts are optional to view and respond to. In this study, Active alerts were used only for medication change recommendations during order entry, while Passive alerts were used for either medication change recommendations when the patient is currently on a relevant medication, or for educational purposes when the patient is not currently on a relevant medication.

*Scenario Development*

A total of 24 different clinical scenarios were developed, using fictional patient information. Each scenario was based on common clinical indications for use of clopidogrel, simvastatin, and warfarin. One scenario was developed for each of the eleven new alert designs, with an additional scenario for warfarin. The additional scenario for warfarin would allow participants multiple exposures to warfarin alerts under different clinical conditions. This resulted in a total of twelve unique scenarios. Analogous scenarios were then developed and paired with the original designs, resulting in a total of 24 testing scenarios. Patient demographics and scenario wording were altered slightly to avoid clear repetition and immediate recognition by participants.

The fictitious patient information for each scenario was based on typical patient demographics for users of these particular medications. Scenarios were evenly split between male and female. Height, weight, and smoking status were based on national averages. Ages were chosen to be clinically reasonable according to national averages for the clinical indication and medication relevant to that scenario.

In addition to patient demographics, each scenario had an associated initial medication (clopidogrel/simvastatin/warfarin), a recommended alternative (prasugrel/atorvastatin/reduced warfarin dose), and an expected action (accept alert/dismiss alert).

All scenarios (including the expected action) were independently reviewed for clinical validity by two MDs before being presented to participants (JBS and one non-author). Reviewers concurred and neither recommended any changes to the scenarios, as written.

*Simulation Development*

Interactive simulations were developed as a web application, using HTML, JavaScript, and CSS. The user interface was created in an image manipulation program and was based on a series of screenshots from NM's actual EHR implementation (Epic Systems Corporation, Verona, WI). All simulated alerts were interactive, with functional hyperlinks, educational materials, and buttons. However, most workflow aspects outside of the alerts were removed or abbreviated for simplicity.  In particular, for scenarios using the original designs, the workflow for ordering after-visit summary materials and for updating a medication were greatly shortened.

Figure 2 shows an example of a simulated workflow using one of the new alert designs.

<u>Figure 2 – Example Simulation</u>

*Screenshot of a simulated EHR sequence in a web application built for user testing of new pharmacogenomic clinical decision support alert designs. Here, the user has attempted to order 75mg of clopidogrel when the patient is a poor metabolizer.*

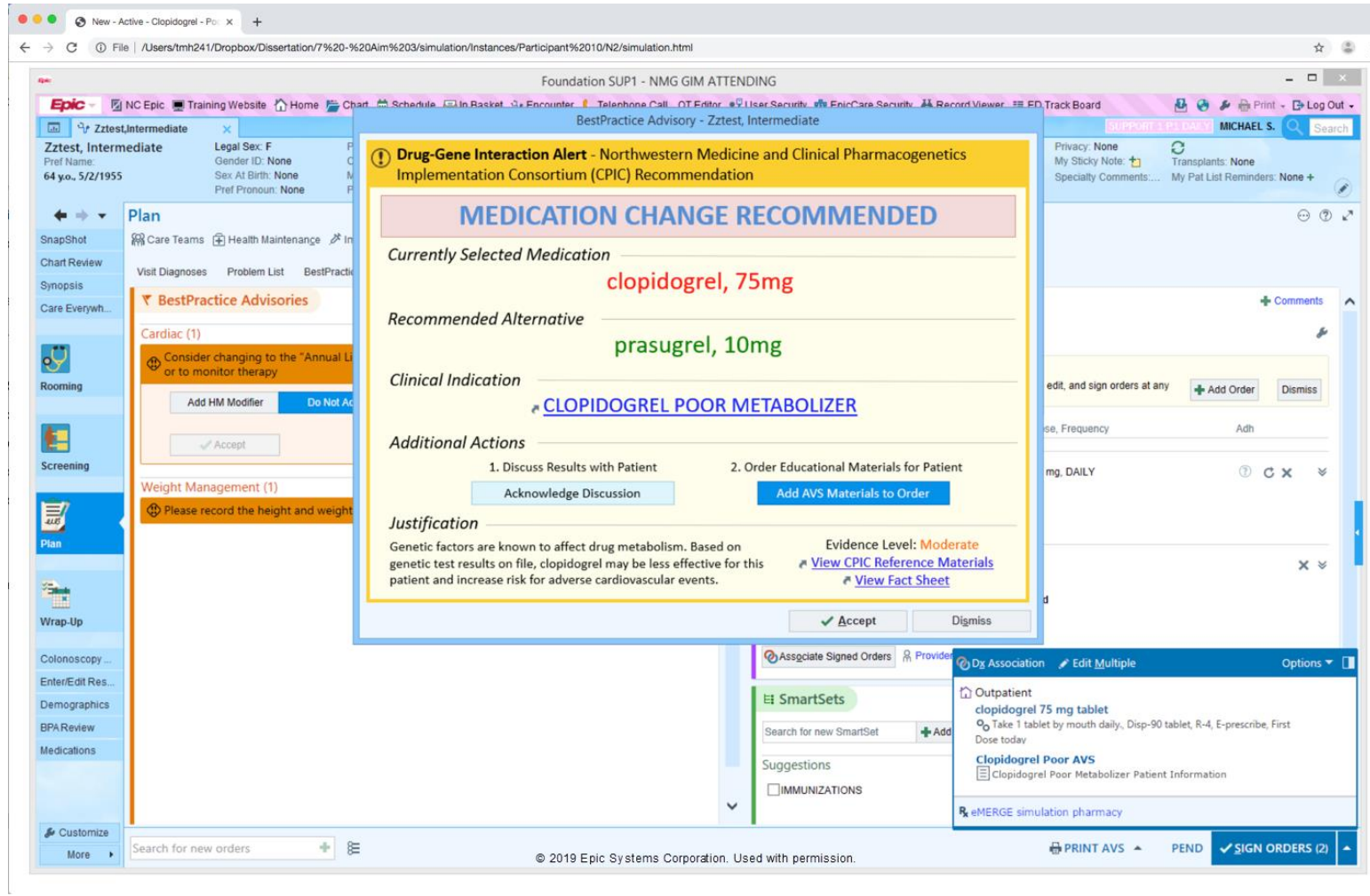

Study Execution

Study participants were recruited from NM's General Internal Medicine (GIM) and Cardiology departments. Participants were approached via e-mail, based on their participation in previous eMERGE studies or through snowball sampling through referrals from other participants. All participants were compensated with a $100 gift card for their participation.

All testing sessions were conducted via think-aloud protocol, either in person or via virtual conference, with the first author present. Participants were encouraged to share their thought processes while executing each of the simulations in order to benefit later review and analysis. Participants were permitted to ask questions about the simulations as they went through their testing session. All sessions were recorded via a virtual conferencing application and were later exported to standalone video. In-person participants were provided a mouse and laptop with the testing application pre-loaded and configured. Virtual conference participants used their own PCs and were provided a copy of the web application ahead of time to run locally.

To mitigate potential learning effects, participants were evenly divided into "Original First" and "New First" conditions. The scenarios and associated simulations were identical for all participants, but the simulation order varied, as follows: The Original First group saw the twelve original designs first, followed by the twelve new designs. The New First group saw the twelve new designs first, followed by the twelve original designs. The order of the scenarios within each "bundle" was randomized for each participant. There was no "washout period" between the bundles of scenarios, as the participants proceeded immediately from the first group to the second group.  Participants were assigned to their respective groups on an alternating basis.

The overall study flow was identical for each participant, outside of the simulation order. Each session started with the participant consenting to the study, followed by five verbal demographics questions assessing their experience, education, and familiarity and comfort with PGx.  Participants were then provided with the web application for user testing.  The web application began with an instruction screen, from which participants clicked a link to begin the study.  Participants then saw a Scenario Description screen with a short paragraph detailing the clinical scenario and a link to begin the simulation.  They would then execute the simulation and complete it by clicking a "Sign Visit" or "Sign Orders" button.  This was followed by a short screen asking them to rate their confidence in their choice on a 0-10 Likert scale and, if they ignored or dismissed an alert, asking them to verbally explain their rationale to the interviewer. Participants would then click a link to view the next Scenario Description screen and repeated this process for all 24 simulations.  After the final post-simulation screen, participants were taken to a wrap-up questionnaire where they answered five Likert-scale questions to assess their preferences between the original and new designs, and three open-ended questions to determine any recommendations they may have for further improving the alert designs.  All questions asked for opinions on "Type 1" or "Type 2" alerts.  Type 1 and Type 2 were alternated for each participant, depending on whether they saw the original designs or new designs first.

Analysis

Analysis was divided into four key metrics: Response, Satisfaction, Speed, and Confidence, each with a different means of assessment.

In concordance with prior work by this research group,[12,13] Response was defined in two ways: Alert Response and Clinical Response.  Alert Response was the response to the alert itself in the simulated EHR and could be either Accept (i.e., the user clicked the "Accept" button on the alert) or Ignore (i.e., the user clicked the "Dismiss" button on the alert or proceeded with no interaction on the alert).  Clinical Response was the clinical action taken and could be either Followed (i.e., the user ordered the recommended alternative medication) or Not Followed (i.e., the user did not order the recommended alternative medication).  The expected and actual responses were were compared for both the Alert Response and Clinical Response.  Each participant viewed many simulations, so each individual action is not independent.  Therefore, we assigned individuals to a High Compliance or Low Compliance category for analysis via McNemar's test.  Participants were assigned to the Low Compliance category if they performed the expected action less than 80% of the time and to the High Compliance category if they performed the expected action at least 80% of the time.  Discordance between Alert Response and Clinical Response (i.e., prescribing actions that do not match the action a clinician took on an alert for a particular scenario) was similarly tested.

Satisfaction was measured based on five Likert-scale questions asked at the end of each testing session.  Additionally, the mean of the five satisfaction scores was calculated to derive a single Overall Satisfaction score.  For analysis, all satisfaction scales were normalized to refer to "original design" and "new design" alerts instead of "Type 1" and "Type 2."  Likert-scales were a five-point scale that ranged from "Significantly prefer original design" to "Significantly prefer new design" and were assigned values from -2 to +2, with Neutral being 0.

Speed was determined by the number seconds elapsed from the time the participant clicked the "Begin Simulation" link for each scenario to the time they clicked the "Sign Visit" or "Sign Orders" button to complete the scenario. This value was determined through a frame-by-frame examination of the session recordings by the first author.

Confidence was determined by a combination of the individual "0-10" confidence scores participants provided at the end of each scenario, as well as an overall confidence question asked during the wrap-up questionnaire phase.

**Results**

Basic Demographics

Twelve clinicians agreed to take part in this study, including eleven MDs and one PA. Six participants were board certified in internal medicine only and six were board certified in cardiology-related specialties. Participants had an average of 18 years of experience and spent 13.5 hours per week seeing patients. When asked about their comfort with genetics and pharmacogenetics, four participants expressed low comfort levels, six expressed fair or moderate comfort levels, and two expressed high comfort levels. Participants generally reported infrequent use of genetic data in their practice, with only one participant saying they used genetic data often. This is consistent with responses seen in previous publications.[12,13]

A technical issue compromised the measurements for one participant during the study. This participant viewed five of the new alert designs and nine of the original designs, instead of twelve of each. This participant's results are excluded in the individual response and time-based analyses but are included for the overall attitude-based analyses. One participant skipped a single scenario in the study due to a technical error and viewed only eleven of the twelve original designs. This participant's results are fully included in all analyses.

<u>Response</u>

For the Alert Response metric, there was little evidence that the new alert designs affected clinicians' choices to Accept or Ignore alerts. There was no detectable difference in compliance rates between the original designs and new designs. Of the 131 scenarios where participants saw the original design, they performed the expected Accept or Ignore action 91 times (69.5%). Of the 132 scenarios where participants saw the new alert design, they performed the expected action 96 times (72.7%). For statistical analysis, participants were labeled as High Compliance if they performed the expected action at least 80% of the time. Compliance statuses are reported in Table 3. The compliance rates with the original and new designs were identical. (McNemar's Test; $H_0$: $p_b = p_c$; p=1)

Table 3 – Alert Response Compliance Rates (# of Participants)

| | | Original Design | | |
| --- | --- | --- | --- | --- |
| | | High Compliance | Low Compliance | **Total** |
| New Design | High Compliance | 3 | 2 | **5** |
| | Low Compliance | 2 | 4 | **6** |
| | **Total** | **5** | **6** | **11** |

For the Clinical Response metric, there was a trend towards better adherence to alert recommendations, but this study lacked the power to demonstrate a statistically significant difference. Of the 109 scenarios where participants saw the original design and were expected to order a medication, they ordered the expected medication 72 times (66.1%). Of the 110 scenarios where participants saw the new alert design and were expected to order a medication, they ordered the expected medication 84 times (76.4%). Again, participants were labeled as High Compliance if they performed the expected action at least 80% of the time. Compliance statuses are reported in Table 4. Despite lacking statistical power to demonstrate significant difference, compliance rates trended towards improved compliance rates with the new designs. (McNemar's Test; $H_0$: $p_b = p_c$; p=0.13) Alternative thresholds for the compliance labels did not alter this conclusion.

Table 4 – Clinical Response Compliance Rates (# of Participants)

|  |  | Original Design | | |
|---|---|---|---|---|
|  |  | High Compliance | Low Compliance | **Total** |
| New Design | High Compliance | 3 | 4 | **7** |
|  | Low Compliance | 0 | 4 | **4** |
|  | **Total** | **3** | **8** | **11** |

Discordance between Alert Response and Clinical Response was present with the original designs (i.e., clinicians did not always perform a clinical action that matches their response to an alert) but eliminated with the new designs. Of the 109 scenarios where participants saw the original design and were expected to order a medication, they performed discordant actions 27 times (24.8%). Of the 110 scenarios where participants saw the new alert design and were expected to order a medication, they performed discordant actions 0 times (0%). Discordance rates are reported in Table 5. Discordance between Alert Response and Clinical Response was eliminated with the new designs. (McNemar's Test; $H_0$: $p_b = p_c$; p=0.004)

Table 5 – Alert Response vs. Clinical Response Discordance Rates

| | | Original Design | | |
|---|---|---|---|---|
| | | 0% Discordance | >0% Discordance | **Total** |
| New Design | 0% Discordance | 1 | 10 | **11** |
| | >0% Discordance | 0 | 0 | **0** |
| | **Total** | **1** | **10** | **11** |

Satisfaction

Participants showed a strong preference for the new alert designs. On the -2 to +2 scale (significantly prefer original designs to significantly prefer new designs), the mean Overall Satisfaction score was 1.37 (one sample, one-tailed t-test; $H_0$: $\mu$=0; p=0.0000001). Overall Satisfaction does not appear to be affected by the New First or Original First condition into which the participant was placed. Those in the Original First condition had an average Overall Satisfaction score of 1.30, while those in the New First condition had an average Overall Satisfaction score of 1.43. This difference was not statistically significant (two sample t-test; $H_0$: $\mu_1$=$\mu_2$; p=0.61).

Responses to the individual satisfaction questions leaned heavily towards a preference for the new designs. Only one participant, on one question, preferred the original designs to the new designs. All other responses expressed preference for the new designs or were neutral. Figure 3 shows how each participant answered each of the five questions and Figure 4 shows the preference distributions for each question. All five individual questions showed preference for the new designs, via one-sample Wilcoxon Signed Rank analysis, both without and with Bonferroni correction for multiple tests. Results on the individual questions are summarized in Table 6.

# Figure 3 – PGx CDS Version Preferences

*Preferences expressed by twelve participants comparing Original and New pharmacogenomic clinical decision support alert designs.*

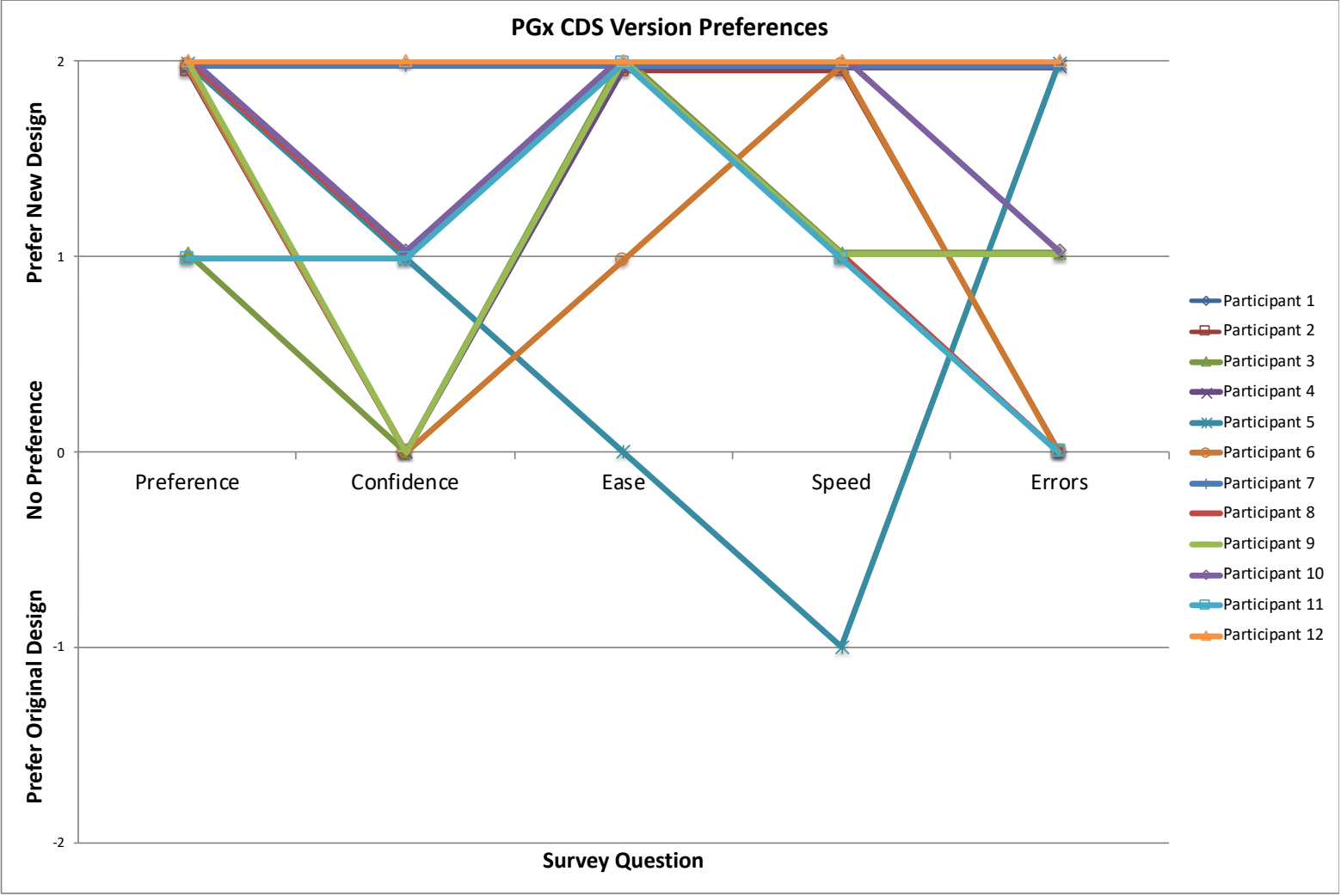

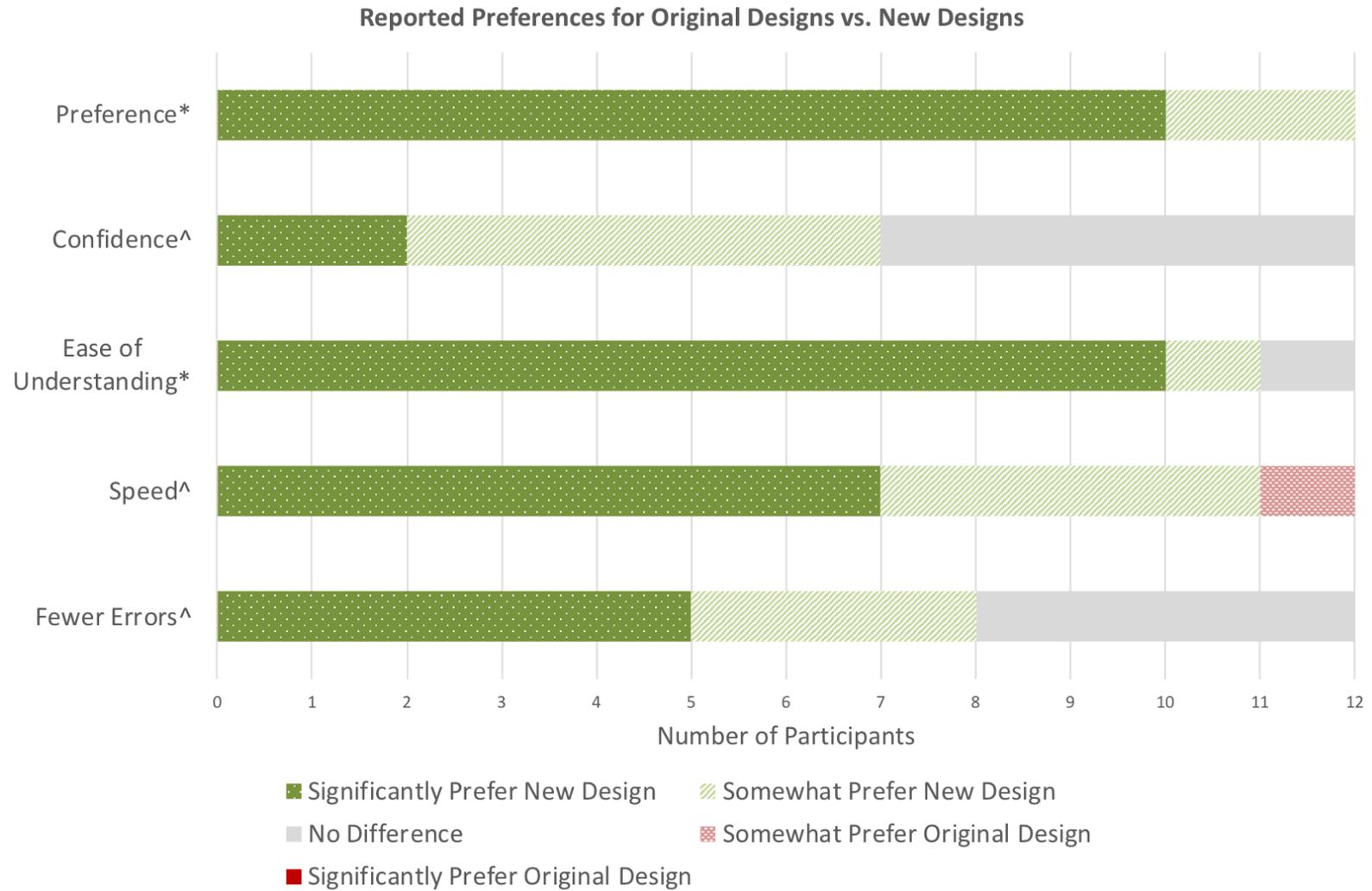

**Reported Preferences for Original Designs vs. New Designs**

*: μ>0; Wilcoxon Signed Rank Test: P<0.001; Bonferroni corrected: P<0.01
^: μ>0; Wilcoxon Signed Rank Test: P<0.01; Bonferroni corrected: P<0.05

Speed

Participants completed tasks significantly faster with the new alert designs than with the original alert designs. The average time taken to complete a single simulation was 61 seconds (SE: 7.7) with the original designs and 40 seconds (SE: 4.3) with the new designs, for a difference of 21 seconds (paired two sample for means, one-tail t-test; $H_0$: $\mu_1 \leq \mu_2$; p=0.009; 99% CI: $0.5 < \mu_2 - \mu_1 < 42.4$).

Learning Curve

Most of the difference in task time comes from the earliest tasks in each sequence, indicating a likely learning curve for the original designs. Participants who saw the original designs first were much slower at earlier simulations than they were at later ones. Conversely, participants who saw the new alert designs first showed similar task times from start to finish. Figures 5 and 6 show the task times for each participant, separated by the Original First and New First conditions.

<u>Figure 5 – Task Times (Original First)</u>

*Simulation task times for each participant who saw Original designs for tasks 1-12 and New designs for tasks 13-24. Results show a decrease in average task time as the simulations progress. (Time: m:ss)*

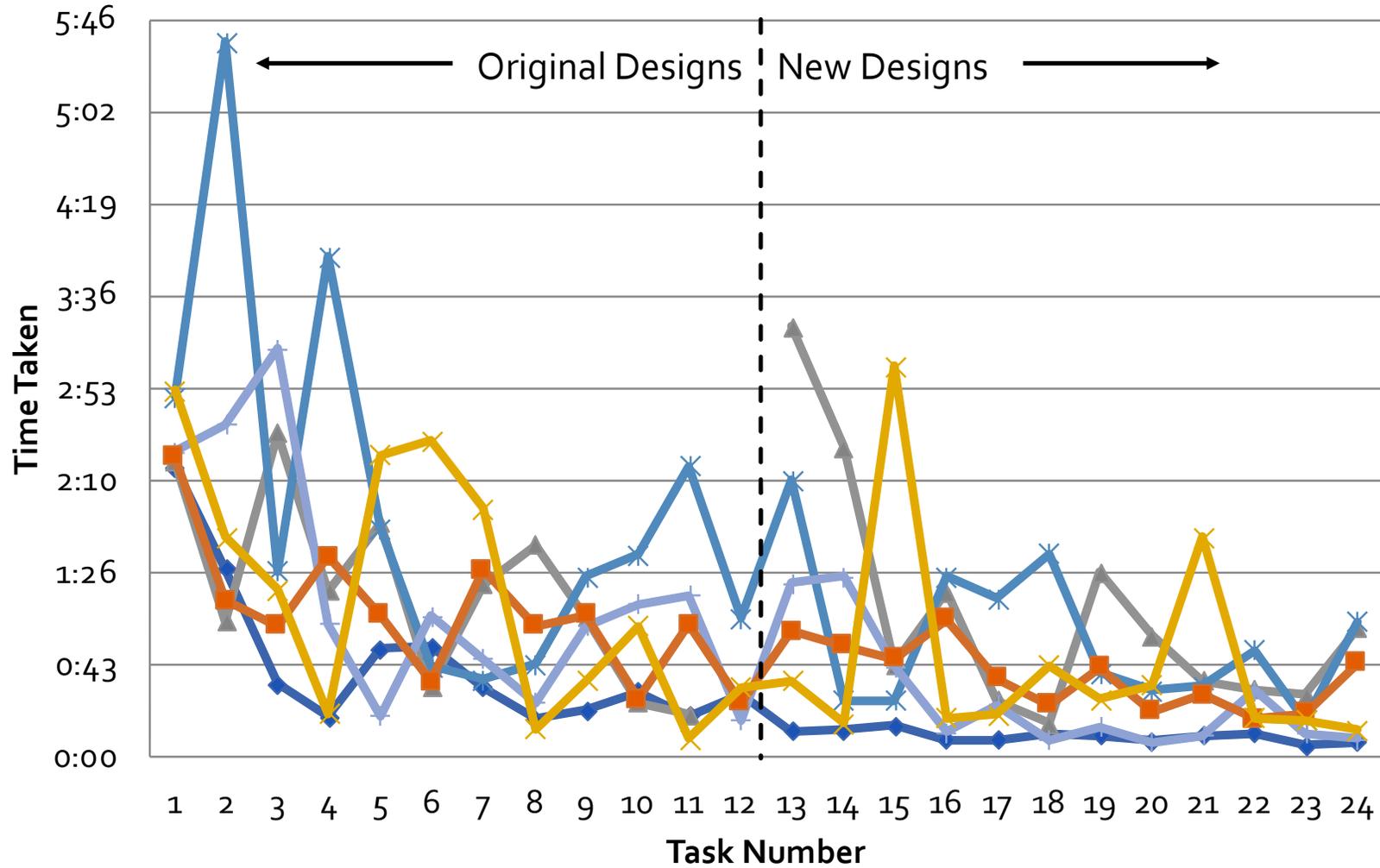

*Simulation task times for each participant who saw New designs for tasks 1-12 and Original designs for tasks 13-24. Results show a consistent task time as the simulations progress. (Time: m:ss)*

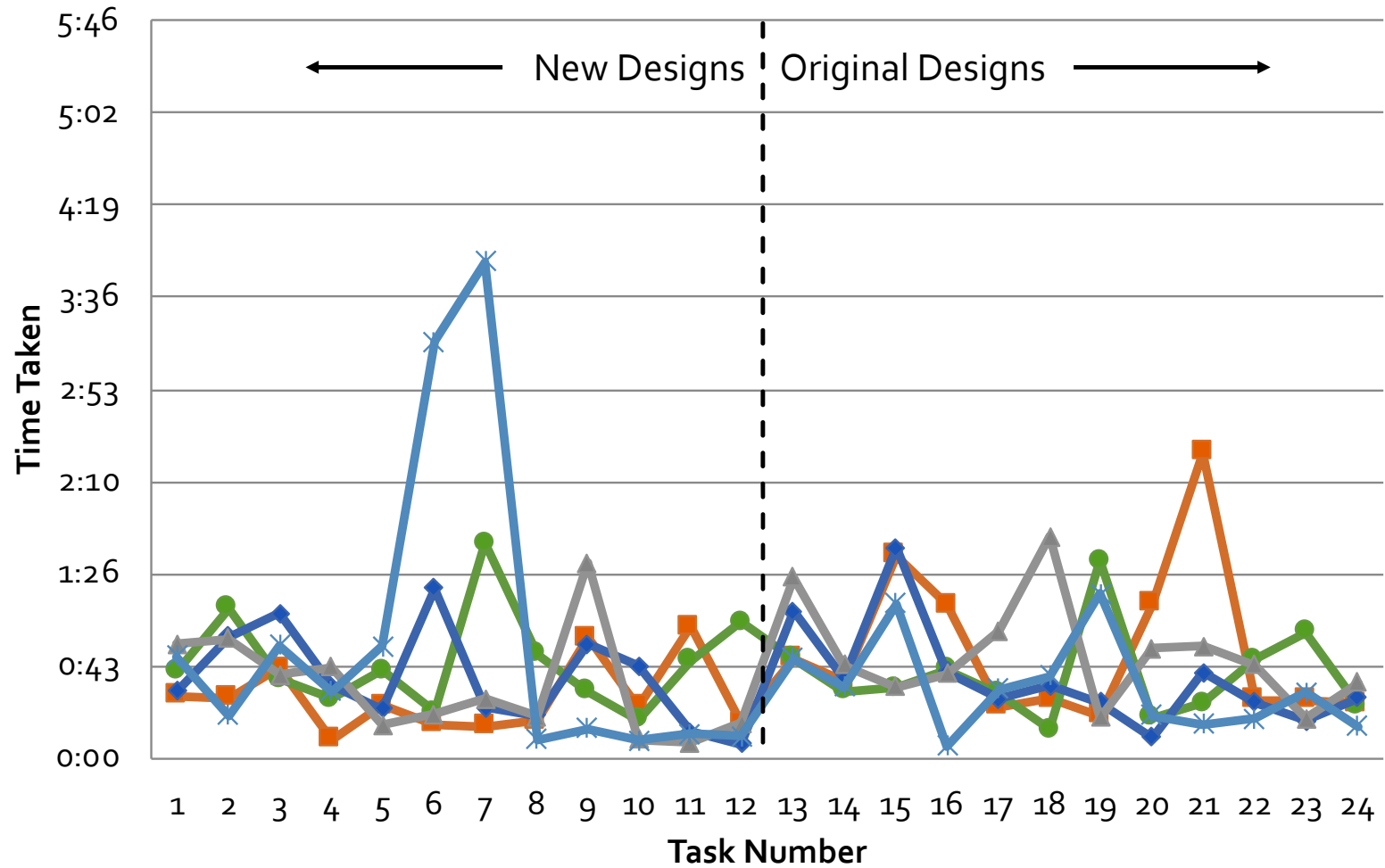

Because of the randomized nature of the simulation order, direct one-to-one comparisons for each task was not feasible. So, to evaluate the possible learning curve, we examined the mean task times in three separate buckets: First Four, Second Four, and Third Four. Table 6 shows that there is a significant time difference in the first four tasks for the Original First group over the New First group. The difference in mean times for the Second Four and Third Four buckets is much smaller and not statistically significant. Furthermore, there was no difference in average task time between the original designs and new designs when seen second, at 42 and 39 seconds, respectively (two sample, one-tail t-test; $H_0$: $\mu_1 \leq \mu_2$; p=0.34). Taken all together, this is evidence that the time differences are due to a substantial learning curve with the original designs that is not present with the new designs, and that the learning curve exists within the first four exposures.

Table 6 – Grouped Averages, Mean Time Difference, Original First vs. New First

| Scenarios | Original First (M:SS) | New First (M:SS) | Difference (M:SS) | P-value (one-tailed t-test) | Bonferroni Corrected |
|---|---|---|---|---|---|
| First 4 | 1:58 | 0:41 | 1:17 | 0.00556274 | 0.016688219 |
| Second 4 | 1:04 | 0:50 | 0:14 | 0.273955594 | 0.821866781 |
| Third 4 | 0:49 | 0:30 | 0:19 | 0.068835658 | 0.206506973 |

Confidence

Participants reported a small but statistically significant increase in confidence levels with the new alert designs. They reported an average confidence score of 7.81 when acting on an alert with an original design, compared to 8.16 when acting on an alert with a new design, for an increase of 0.35 (Paired two sample for means, one-tail, t-test; $H_0$: $\mu_1 \leq \mu_2$; p<.05).

**Discussion**

This study successfully demonstrates that socio-technical techniques can improve PGx CDS alert designs. Participants in this study showed greater satisfaction, lower task times, greater confidence, and improved concordance of actions when interacting with revised alerts, with no evidence of any deleterious effects on accuracy.

Though we were unable to demonstrate that the new designs lead to greater accuracy, we did find a positive trend in Clinical Response that may be significant with greater statistical power. Additionally, clinician behavior was more concordant between Alert Response and Clinical Response. This is likely due to a workflow change in the new designs. The new designs automatically update the user's order with the recommended medication when they accept an alert (a reflection of the "Be Specific and Actionable" design principle), while the original designs require users to manually update their order. This manual process introduces the opportunity for divergence between the Alert Response and Clinical Response actions, which could cause confusion for users and complications for system evaluators.

The increase in confidence we observed from individual scenario scores is further supported by the results from the end-of-study questionnaire, where participants reported being more confident overall with the new designs. However, that result was the weakest of the five end-of-study questions, which aligns with the relatively modest increase in confidence that we found.

The Alert Response and Clinical Response rates in this study were significantly higher than rates found in live clinical data in a previous study at NM, where clinicians accepted the original-style alerts 42.5% of the time and ordered an alternative medication 22.5% of the time.[12] These rates are substantially lower than in this study, where clinicians accepted the original designs 69.5% of the time and ordered an alternative medication 66.1% of the time. There are several possible reasons for this result, including that the simulations in this study were not time-bound, so participants were able to take the time they needed to review alerts and make a choice in which they felt confident. In live clinical settings, users may be reluctant to take the time to review the unfamiliar alerts and change their behavior, particularly given the substantial learning curve we found for the original designs. We do not see any reason to believe this would bias one condition over the other in our study. Also, given the reduced learning curve and task times with the new designs, clinicians may be even more likely to follow the new designs in a time-constrained setting, leaving our overall conclusions unchanged. Additionally, the Hawthorne Effect[26] may be present in this study, as participants were fully aware that they were being observed. But again, we see no reason to believe this would bias one condition over the other, and therefore do not believe our conclusions are affected.

The present study adds to a small but growing base of literature examining the effectiveness of PGx CDS, of which some previous studies have also utilized socio-technical approaches. Our results dovetail with previous studies from the University of Washington. Devine, et al. found that physicians expressed a desire for PGx CDS to "provide succinct, relevant guidelines and dosing recommendations of phenotypic information from credible and trustworthy sources; any more information was overwhelming."[27] Here, we find that such designs do, in fact, improve physician satisfaction with PGx CDS. Subsequently, Overby et al. performed an evaluation of a PGx CDS prototype and found that PGx CDS lowered physician confidence in their decision-making, but significantly altered their dosing decisions.[28] Our results contradict those findings, as we found an increase in confidence but little change in decision making. It is likely that differences in design, workflow, or physician characteristics contributed to these divergent results, which emphasizes the need for further research to fully understand how such factors affect clinician response to PGx CDS. Our findings are congruent with their conclusion that effective PGx CDS is "likely to be realized through continued focus on content, content delivery, and tailoring to physician characteristics."[28]

Other reports of PGx CDS effectiveness include a study by O'Donnell, et al. that found PGx CDS was effective in altering prescriber behavior.[14] Their Genomic Prescribing System differs dramatically from the designs at NM, relying on a standalone web application using "traffic light" style alerts to provide recommendations, as opposed to interruptive alerts integrated into the EHR. Similarly, Bell, et al. found 95% compliance rates with PGx CDS in treating pediatric lymphoma and leukemia patients.[16] Conversely, St. Sauver, et al. found only 30% compliance with PGx CDS, and that "clinicians felt that the alerts were confusing, irritating, frustrating, or that it was difficult to find additional information."[29] Results from an eMERGE-wide pilot study of clinician response to PGx CDS were heavily mixed, with alert adherence rates ranging from 21% to 73%, depending on the site.[13] Melton, et al. found that incorporating user feedback during development of warfarin-based CDS increased acceptance of recommendations, but did not alter overall satisfaction with system prototypes.[30] The present study fits the broader body of literature in that it underscores the lack of consensus around optimal PGx CDS design and clinician response, and provides promising direction for those that wish to avoid designs that would frustrate or irritate users.

This study does not seek to evaluate the appropriateness and value of PGx CDS in general, nor does it seek to recommend particular clinical scenarios where PGx CDS may be most beneficial. Instead, this study focuses exclusively on improving alert designs so that they may be more effective in any particular scenario where an individual healthcare organization has deemed PGx CDS necessary and/or beneficial.

The simulated nature of the study does not perfectly mimic an actual clinical setting with all the time pressures, potential distractions, and emotional involvement of an actual patient that such a setting would entail. We instructed clinicians to work quickly and act as they would with a real patient, but the extent to which they were able to accomplish this is unclear.

The times collected for the speed metric are not likely to reflect actual clinical times because of the think-aloud protocol the study employed. However, there is little reason to believe that the time required to think aloud would change the conclusion that the new designs are faster for clinicians. Several articles in the literature discuss how think-aloud protocols affect task times and accuracy, but we were unable to find any case where such protocols would bias one study condition over another when both conditions include a think-aloud component.[31,32] Additionally, as described in the Materials and Methods section, the simulations for the original designs were significantly shortened to reduce complexity in the web application. The simulations for the new designs were not shortened, a fact that further strengthens our conclusions. Moreover, most of the time difference was due to a large learning curve with the original designs, where early exposures took clinicians much longer to respond. Given the low frequency of interaction that clinicians have with PGx CDS in real-world clinical settings,[12] we expect that the earlier, slower times in this study more closely reflect real-world interactions.

With these limitations in mind, we conclude that the new designs are faster, with a reduced learning curve, but we make no claim as to the magnitude of the change in real-world task time.

This study is further limited by its sampling methods. Participants were not randomly sampled, but instead were clinicians that had worked with PGx before, had participated in a prior eMERGE-PGx study, or who showed interest in the project via professional networking. Furthermore, participation was limited to GIM and Cardiology departments at a single large, academic medical center. All of these factors make it possible that participants were more informed about PGx in general, and as such, the results may not generalize to other clinical populations. However, the low to moderate comfort level with, and infrequent use of, genetics that participants reported may indeed reflect the attitudes of a typical clinician.

Future work can build upon this study in a number of ways. We recorded the user testing sessions and asked participants to think aloud during their simulations, which could provide additional fruitful data. The final wrap-up survey also asked open-ended questions that encouraged participants to suggest improvements to the new designs. These qualitative analyses are left for a future study. Anecdotally, there are minor tweaks that could lead to even greater satisfaction with the new designs – such as a "Preview" link for the patient educational materials and renaming the "Accept" button to "Order [alternative]" to further clarify the workflow.

Most importantly, this successful pilot demonstration should be followed up with a real-world implementation and evaluation. To the extent that it is technically feasible, the alert designs in this study should be translated to live alerts in an active EHR and subjected to similar evaluations as our previous studies of the original alert designs.[12] Iteration and continual improvement are critical aspects of socio-technical design.

**Conclusion**

This study validates the PGx CDS design principles we previously proposed and provides strong evidence that socio-technical design approaches lead to better results with PGx CDS than design by committee.  The updated designs tested in this study led to more satisfied, faster, and more confident clinicians who were at least as accurate in their decision making as with prior, committee-developed designs.  These results support previous findings that, for PGx, clinicians prefer brief, actionable alerts that contain specific recommendations based on interpreted phenotypes.  As a next step, these alert designs should be implemented in a live clinical setting to confirm that they lead to similar real-world results.